\documentclass[a4paper,pre,twocolumn,superscriptaddress,showkeys,floatfix]{revtex4-1}
\usepackage{graphicx}
\usepackage{amsmath}
\usepackage{amsthm}
\usepackage{amssymb}
\usepackage{mathrsfs}

\newcommand{\mean}[1]{\left\langle #1 \right\rangle}   
\newcommand{\var}[1]{\mathrm{var}\!\left( #1 \right)}  
\DeclareMathOperator{\mkurt}{\mathcal K}               
\DeclareMathOperator{\eskew}{\mathcal S}               
\DeclareMathOperator{\ekurt}{\mathcal K}               
\DeclareMathOperator{\erg}{\mathcal U}                 
\DeclareMathOperator{\heat}{\mathcal C}                

\begin{document}

\title{Boundary effects on finite-size scaling for the 5-dimensional
  Ising model}

\date{\today} 

\author{P. H. Lundow} 
\email{per.hakan.lundow@math.umu.se} 


\affiliation{Department of mathematics and mathematical statistics,
  Ume\aa{} University, SE-901 87 Ume\aa, Sweden}

\begin{abstract}
  High-dimensional ($d\ge 5$) Ising systems have mean-field critical
  exponents. However, at the critical temperature the finite-size
  scaling of the susceptibility $\chi$ depends on the boundary
  conditions. A system with periodic boundary conditions then has
  $\chi\propto L^{5/2}$. Deleting the $5L^4$ boundary edges we receive
  a system with free boundary conditions and now $\chi\propto L^2$.
  In the present work we find that deleting the $L^4$ boundary edges
  along just one direction is enough to have the scaling $\chi\propto
  L^2$. It also appears that deleting $L^3$ boundary edges results in
  an intermediate scaling, here estimated to $\chi\propto
  L^{2.275}$. We also study how the energy and magnetisation
  distributions change when deleting boundary edges.
\end{abstract}


\maketitle

\section{Introduction}
The 5-dimensional (5D) Ising model is well-known to have mean-field
critical exponents so that, for example, $\alpha=0$, $\gamma=1$ and
$\nu=1/2$. Assuming the usual finite-size scaling (FSS) rules this
would imply that the susceptibility scales as $\chi\propto
L^{\gamma/\nu}=L^2$ near the critical point $\beta_c$, where $L$ is
the linear order of the system. However, since we are above the upper
critical dimension $4$ this rule breaks down and we find instead
$\chi\propto
L^{d/2}=L^{5/2}$~\cite{brezin:85,binder:85,blote:97,luijten:99}, at
least for periodic (cyclic) boundary conditions (PBC).

Already in Ref.~\cite{rudnick:85} was it suggested on theoretical, if
non-rigorous, grounds that for free boundary conditions (FBC) the rule
$\chi\propto L^2$ holds, for $d\ge 5$. This was ultimately settled in
Ref.~\cite{camia:20} by rigorous means (under some mild assumptions),
after a fruitful
debate~\cite{lundow:11,lundow:14,berche:12,berche:16}. However, this
only concerned the critical point $\beta_c$ and it is still open
whether $\chi\propto L^{5/2}$ for some $L$-dependent point
$\beta(L)$~\cite{young:2014,lundow:16}.

In the present work we investigate the effect of boundary conditions
between FBC and PBC. We thus start with a PBC-system and delete, for
example, all boundary edges along one or more, say $r$,
dimensions. This means we delete $rL^4$ edges and, as we will see,
this is enough to change the scaling behaviour of $\chi$ to that
typical of an FBC-system. Other scenarios for removing boundary edges
are also interesting. Deleting $rL^3$ edges seems to give a scaling
behaviour between that of PBC and FBC, suggesting $\chi\propto
L^{2.275}$. However, deleting only $rL^2$ edges gives a scaling
behaviour indistinguishable from that of PBC.

We will study the behaviour of not only the susceptibility, but also
the distribution shape (kurtosis) and the effects on the energy
distribution (variance, skewness and kurtosis).  However, though we
have collected data for a wide range of system sizes (up to $L=95$)
our investigation only takes place at the critical point.

\section{Definitions and details}
The underlying graph is the $L\times L\times L\times L\times L$ grid
graph on $N=L^5$ vertices but with different cases of boundary
conditions, to be defined below. With each vertex $i$ we associate a
spin $s_i=\pm 1$ and let the Hamiltonian be $\mathscr{H}=\sum_{ij} s_i
s_j$.  The magnetisation of a state $s=(s_1,\ldots,s_N)$ is $M=\sum_i
s_i$, where we sum over the vertices $i$, and the energy is
$E=\sum_{ij} s_is_j$, where we sum over the edges $ij$. Their
normalised forms are denoted $m=M/N$ and $U=E/N$. As usual,
$\mean{\ldots}$ denotes the thermal-equlilibrium mean and
$\var{\ldots}$ the variance. Quantities of interest to us are the
susceptibility $\chi=\mean{M^2}/N$, the internal energy
$\erg=\mean{E}/N$ and the specific heat $\heat=\var{E}/N$ (we ignore
the usual $\beta^2$-factor).  Distribution shape characteristics such
as skewness and kurtosis are also of interest. Since the distribution
of magnetisations is symmetric (hence skewness zero) when no external
field is present, only its kurtosis is considered:
\begin{equation}\label{mkurtdef}
  \mkurt = \frac{\mean{M^4}}{\mean{M^2}^2}
\end{equation}

For the energy distribution its skewness is defined as
\begin{equation}\label{eskewdef}
  \eskew = \frac{\mean{(E-\langle E\rangle)^3}}{\var{E}^{3/2}}
\end{equation}
and its kurtosis as
\begin{equation}\label{ekurtdef}
  \ekurt = \frac{\mean{(E - \langle E\rangle)^4}}{\var{E}^2}
\end{equation}
Hopefully the context will make it clear whether $\mkurt$ is referring
to energy or magnetisation kurtosis.

We have collected data for systems of linear order $L=15$, $19$, $23$,
$31$, $39$, $47$, $55$, $63$, $71$, $79$, $87$ and $95$ using standard
Wolff-cluster updating~\cite{wolff:89}. An expected $N$ spins were
flipped between measurements of energy and magnetisations. All
sampling took place at the critical point
$\beta_c=0.11391498$~\cite{lundow:15}. The number of sample $n_s$ are
between $100000$ and $200000$ for $15\le L\le 63$ and ca $50000$ for
$71\le L\le 95$. However, for PBC we have at least $500000$ samples
for $15\le L\le 79$ and $50000$ for $L=87, 95$. Error bars of the
quantities mentioned above are estimated from bootstrap resampling of
the data.

The vertices of our graph are the integer points in a 5D system,
$V=\{(i_1, \ldots, i_5): 1\le i_1,\ldots, i_5\le L\}$. We let $E$
denote the edges of a PBC-system, which has $5L^5$ edges. There are
two types of edges, the bulk edges and the boundary edges. The bulk
edges are those edges $e=\{i,j\}$, where $i=(i_1,\ldots,i_5)$ and
$j=(j_1,\ldots,j_5)$, such that the Manhattan distance $\sum_{r=1}^5
|i_r-j_r|$ (sum over coordinates) is $1$. The boundary edges belong to
one of five sets, $D_1, \ldots,D_5$. For example, the boundary edges
in $D_1$ are on the form
\begin{equation}
  e=\{(1,i_2,i_3,i_4,i_5),(L,i_2,i_3,i_4,i_5)\},
\end{equation}
where $1\le i_2,i_3,i_4,i_5\le L$. Hence there are $L^4$ such edges in
$D_1$. In general, the set $D_r$ has the $1$ and $L$ at coordinate
$r$, for $1\le r\le 5$.  For convenience we define
$C_r=D_1\cup\ldots\cup D_r$, so that $|C_r|=rL^4$. Thus the FBC-system
only keeps the bulk edges, $E\setminus C_5$, which then has
cardinality $5L^5-5L^4$.

We will now define two subsets of $D_1$. The set of edges on the form
\begin{equation}
  e=\{(1,i_2,i_3,i_4,x),(L,i_2,i_3,i_4,x)\},
\end{equation}
where $1\le x\le r$, is denoted $B_r$ and contains $rL^3$ edges. The
set of edges on the form
\begin{equation}
  e=\{(1,i_2,i_3,x,y),(L,i_2,i_3,x,y)\},
\end{equation}
where $1\le x, y\le r$, is denoted $A_r$ and contains $r^2L^2$ edges.

We have collected measurements of energy and magnetisation for the
following boundary cases: $E$ (or PBC, deleting no edges), $E\setminus
A_r$ (deleting $r^2L^2$ edges, $r=1,2,4$), $E\setminus B_r$ (deleting
$rL^3$ edges, $r=1,2,4$), $E\setminus C_r$ (deleting $rL^4$ edges,
$r=1,2,3,4,5$, where $r=5$ is FBC). However, for short we will usually
refer to these cases as simply PBC, $A_r$, $B_r$ and $C_r$, with $C_5$
synonymous to FBC.

One could of course also consider other interesting cases by letting
the parameter $r$ depend weakly on $L$, but presumably it would
require rather large $L$ to tell a new scaling effect from an extra
correction term. An alternative, in the cases $A_r$ and $B_r$, is to
let the $x$ and $y$ be centred around $L/2$ instead of its current
choice $1\le x,y\le r$.

\section{Scaling of susceptibility}
As we mentioned above we have $\chi\propto L^{5/2}$ for PBC and
$\chi\propto L^2$ for FBC ($C_5$). Here we will see that $\chi\propto
L^{5/2}$ in the $A_r$-case, $\chi\propto L^2$ in the $C_r$-case, but
$\chi\propto L^{2.275}$ for the inbetween case $B_r$.

We begin by taking our data points, $\chi/L^a$ versus $1/L$, and fit a
line $y=c_0+c_1x$ to these, for each boundary case. Testing a range of
exponents $a$ between $2$ and $5/2$ we compute the root-mean-square
(RMS) of the deviation between the fitted line and the points.  A
minimum RMS then indicates the optimal value of $a$.  This ignores any
higher-order corrections but this, as we will see, appears quite
acceptable. Since we know the correct values of $a$ for PBC ($a=2.5$)
and FBC ($a=2$) this also gives us an indication of the error in our
estimate.

\begin{figure}
  \includegraphics[width=0.483\textwidth]{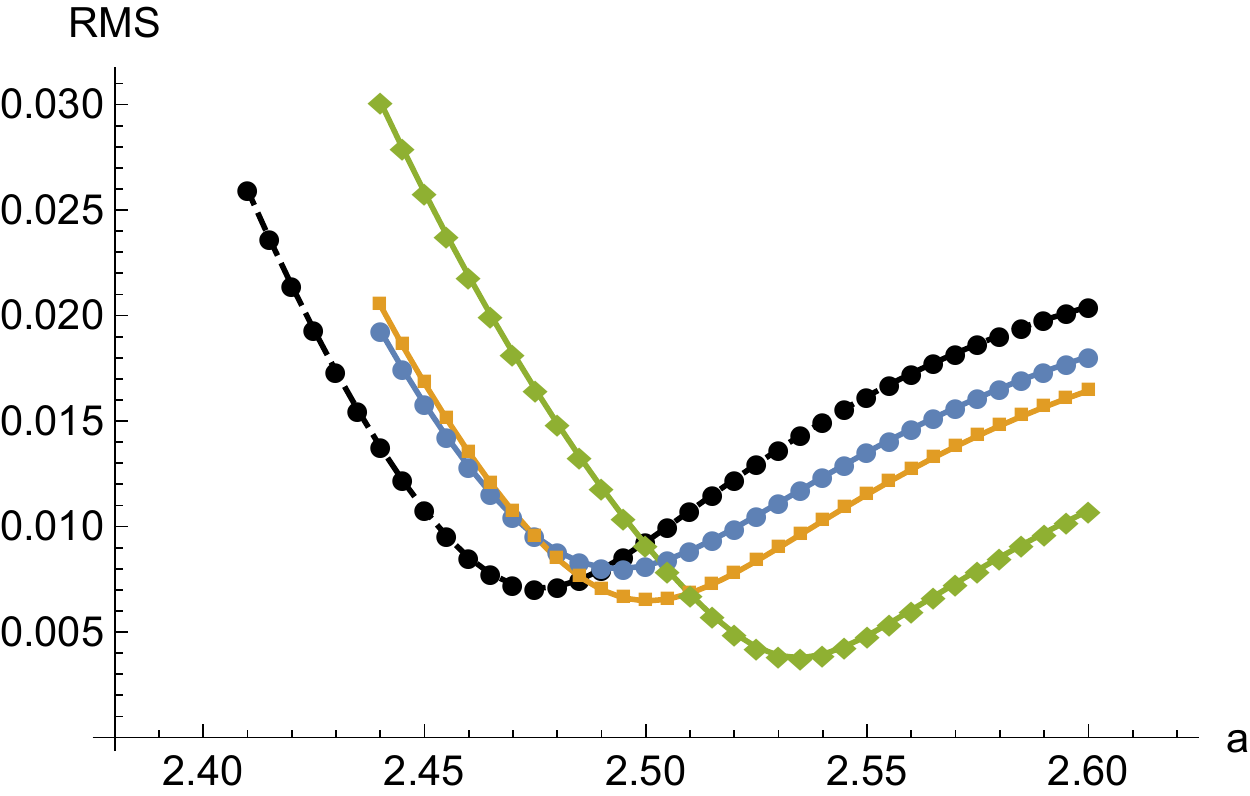}
  \caption{\label{fig:rms02}(Colour on-line) Root-mean-square of
    linear fit to points $(1/L,\chi/L^a)$ plotted versus $a$ for PBC
    (black), $A_1$ (blue), $A_2$ (orange) and $A_4$ (green).  Minimum
    at respectively $2.475$, $2.49$, $2.50$, $2.535$.}
\end{figure}

In Fig.~\ref{fig:rms02} we see that PBC and $A_r$ prefer exponents $a$
near $5/2$, with minimum at $2.475$ (PBC), $2.49$ ($A_1$), $2.50$
($A_2$) and $2.535$ ($A_4$). This gives the estimate
$a=2.500(25)$. Deleting a few of the points in the fit will of course
give slightly different results, but within the error bar.

\begin{figure}
  \includegraphics[width=0.483\textwidth]{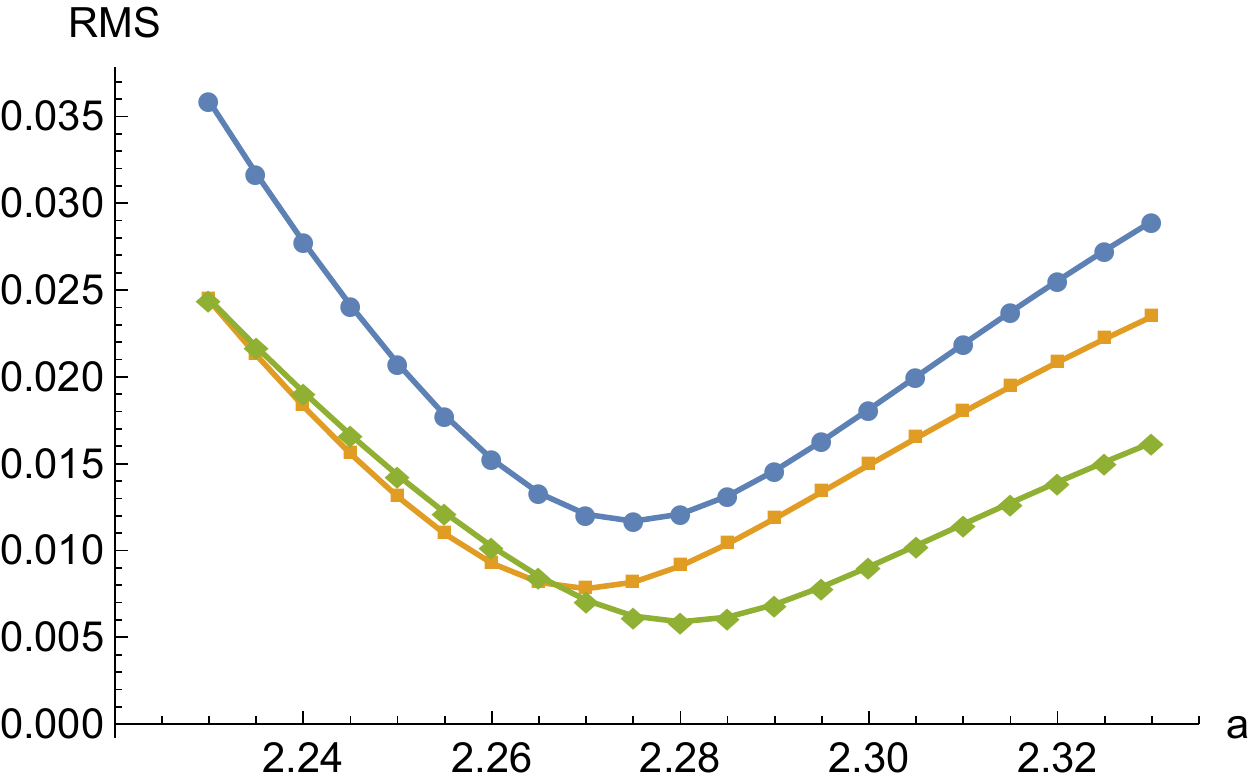}
  \caption{\label{fig:rms3}(Colour on-line) Root-mean-square of linear
    fit to points $(1/L,\chi/L^a)$ plotted versus $a$ for $
    B_1$ (blue), $B_2$ (orange) and $B_4$ (green).  Minimum at
    $2.275$, $2.27$, $2.28$, respectively.}
\end{figure}

Continuing with the $B_r$-case we show RMS versus $a$ in
Fig.~\ref{fig:rms3}. The mean of the minima centre around $2.275(10)$,
where we have used $L\ge 23$ for the fitted line. It would perhaps be
natural to expect the linear fit to favour the mid-point $2.25$
between $2$ and $2.5$ but this is not supported by the present
data. It should be remarked that, the mean value of the minima is
remarkably stable between different point sets but the individual
minima vary between $2.255$ and $2.29$. Hence we suggest that the
correct $a$-value is a little larger than $2.25$.

\begin{figure}
  \includegraphics[width=0.483\textwidth]{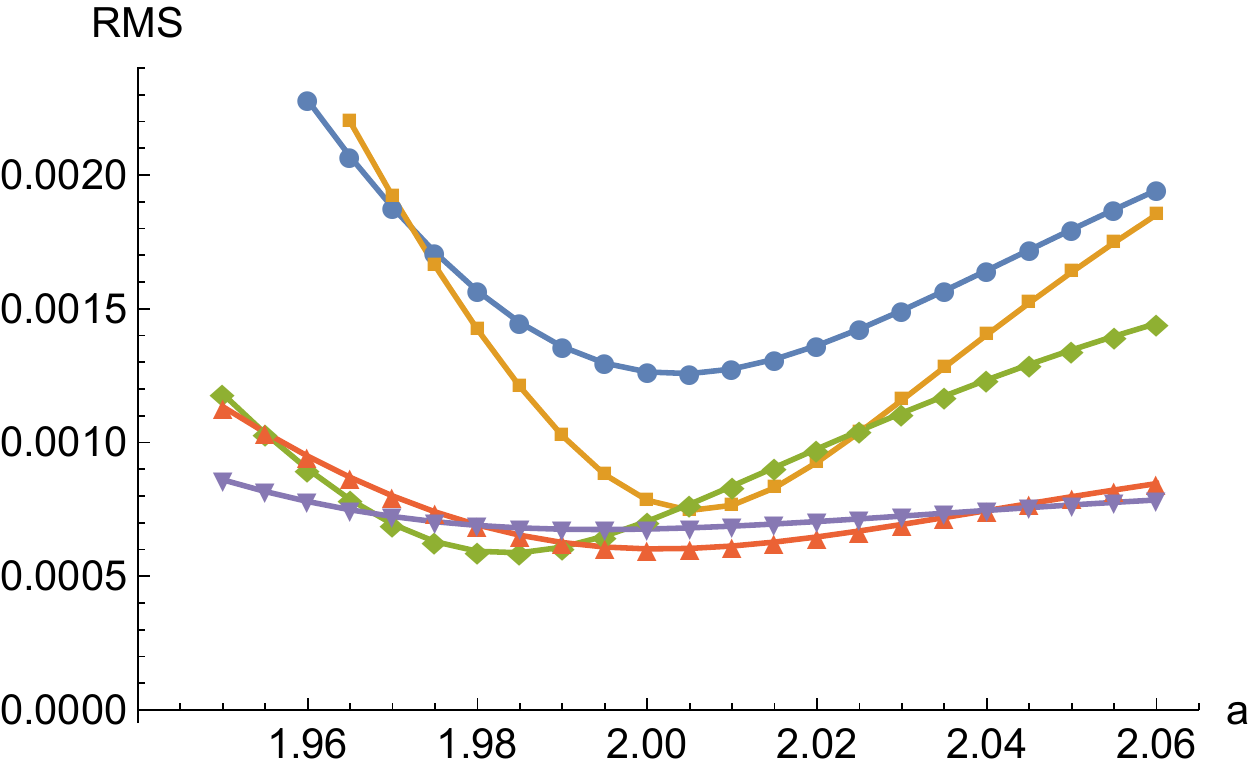}
  \caption{\label{fig:rms4}(Colour on-line) Root-mean-square of linear
    fit to points $(1/L,\chi/L^a)$ plotted versus $a$ for $
    C_1$ (blue, values divided by 3), $C_2$ (orange), $C_3$ (green),
    $C_4$ (red) and $ C_5$ (purple).  Minimum at
    respectively $2.005$, $2.005$, $1.985$, $2.000$, $1.995$.}
\end{figure}

Finally, the case of $C_r$ (thus including FBC) is shown in
Fig.~\ref{fig:rms4}. The RMS-plots clearly prefer an exponent close to
$2$.  With the present linear fit, using $L\ge 19$, we find an average
minimum $2.00(1)$.  Trying different point sets gives almost the same
average but individual minima varies between $1.97$ and $2.015$.

\begin{figure}
  \includegraphics[width=0.483\textwidth]{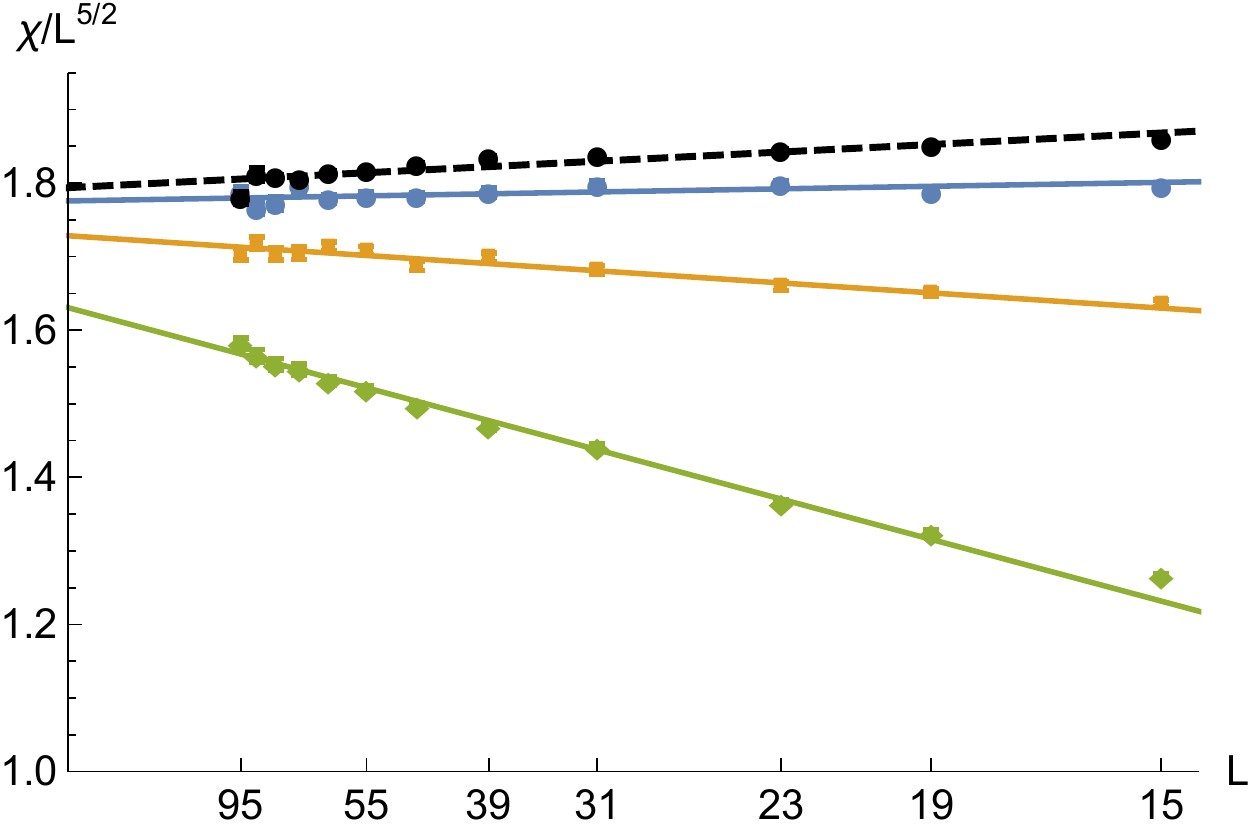}
  \caption{\label{fig:chi2}(Colour on-line) Normalised susceptibility
    $\chi/L^{5/2}$ versus $1/L$ for $L=15$, $19$, $23$, $31$, $39$,
    $47$, $55$, $63$, $71$, $79$, $87$ and $95$. Cases are PBC
    (black), $A_1$ (blue), $A_2$ (orange) and $A_4$ (green).  Fitted
    lines are, respectively, $y=1.794+1.11x$, $y=1.775+0.40x$,
    $y=1.729-1.50x$, $y=1.630-6.0x$ where $x=1/L$.}
\end{figure}

\begin{figure}
  \includegraphics[width=0.483\textwidth]{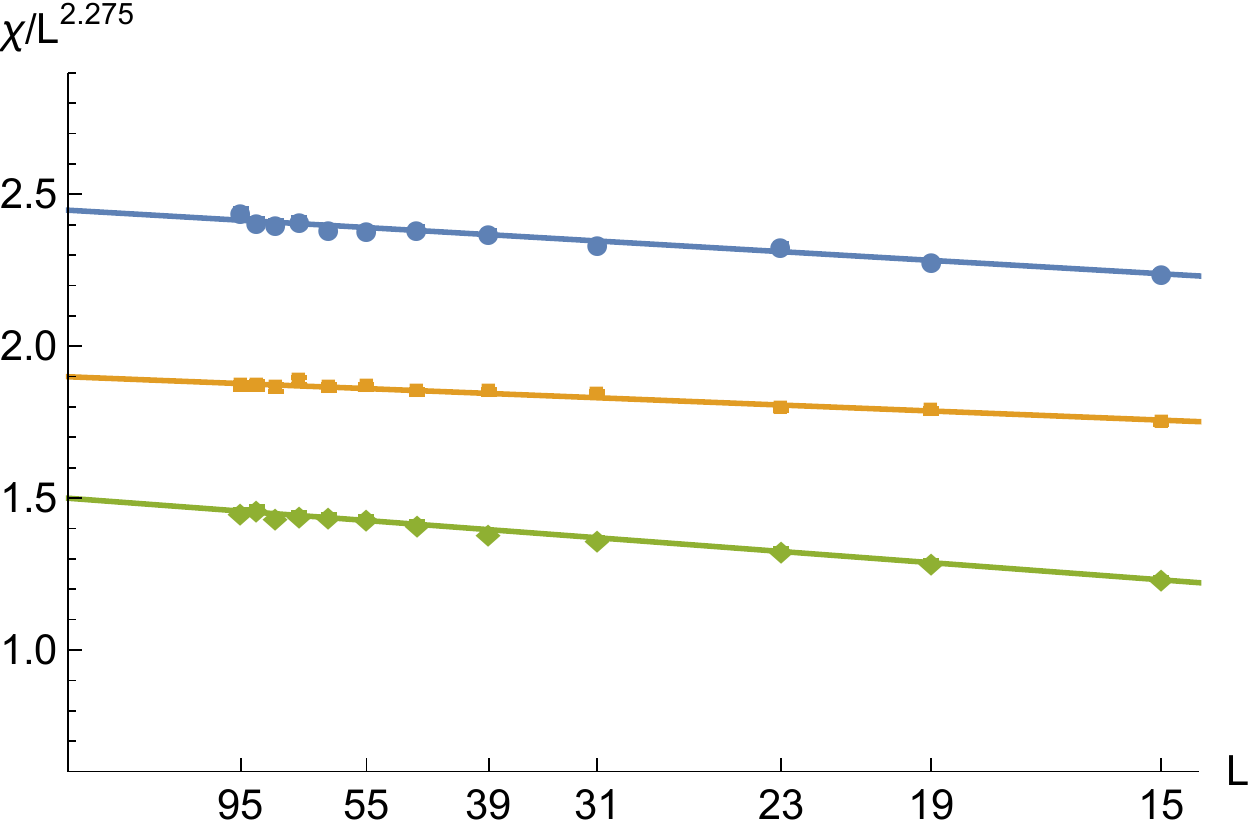}
  \caption{\label{fig:chi3}(Colour on-line) Normalised susceptibility
    $\chi/L^{2.275}$ versus $1/L$ for $L=15$, $19$, $23$, $31$, $39$,
    $47$, $55$, $63$, $71$, $79$, $87$ and $95$. Cases are $
    B_1$ (blue), $B_2$ (orange) and $B_4$ (green).  Fitted lines are,
    respectively, $y=2.45-3.1x$, $y=1.90-2.2x$, $1.50-4.0x$ where
    $x=1/L$.}
\end{figure}

\begin{figure}
  \includegraphics[width=0.483\textwidth]{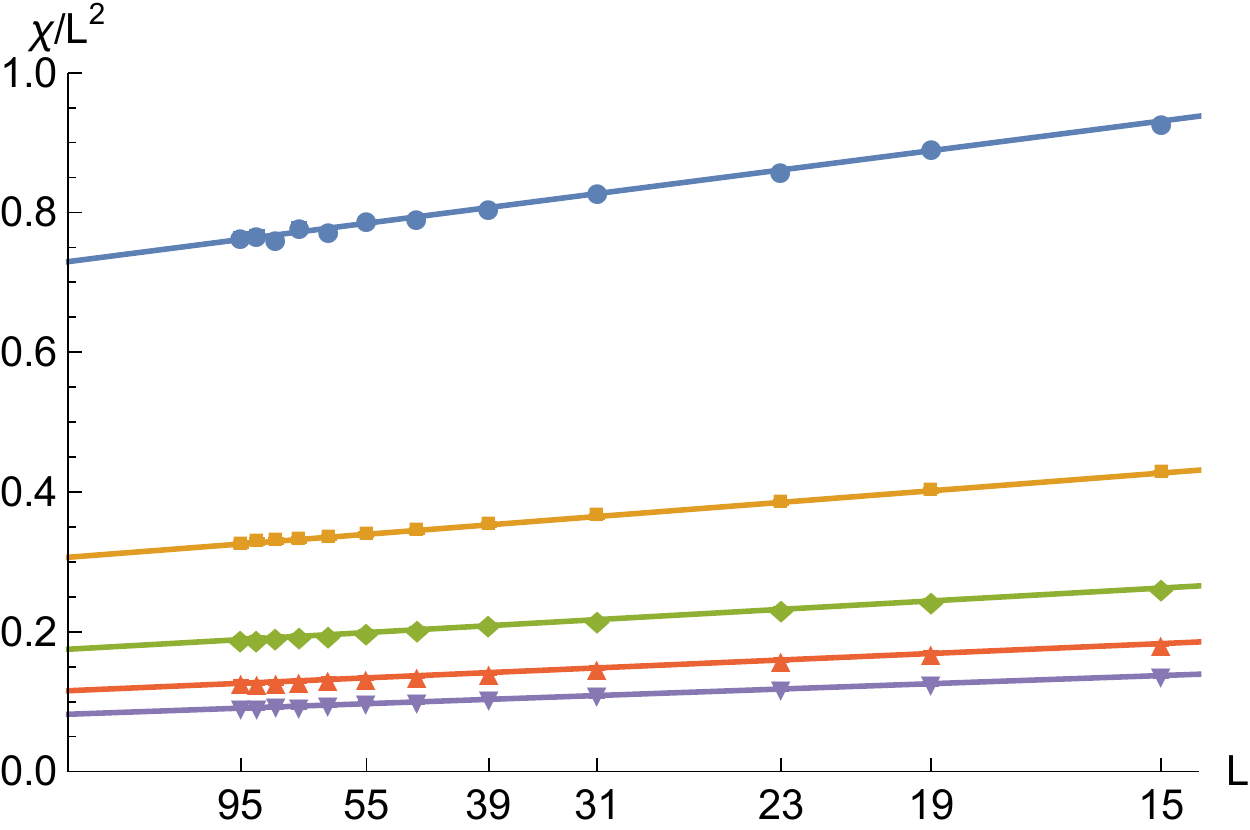}
  \caption{\label{fig:chi4}(Colour on-line) Normalised susceptibility
    $\chi/L^2$ versus $1/L$ for $L=15$, $19$, $23$, $31$, $39$, $47$,
    $55$, $63$, $71$, $79$, $87$ and $95$. Cases are (downwards) $C_1$
    (blue), $C_2$ (orange), $ C_3$ (green), $C_4$ (red) and
    $C_5$ (purple).  Fitted lines are, respectively, $y=0.730+3.0x$,
    $y=0.307+1.8x$, $0.175+1.3x$, $y=0.116+1.0x$, $y=0.082+0.83x$
    where $x=1/L$.}
\end{figure}

\begin{figure}
  \includegraphics[width=0.483\textwidth]{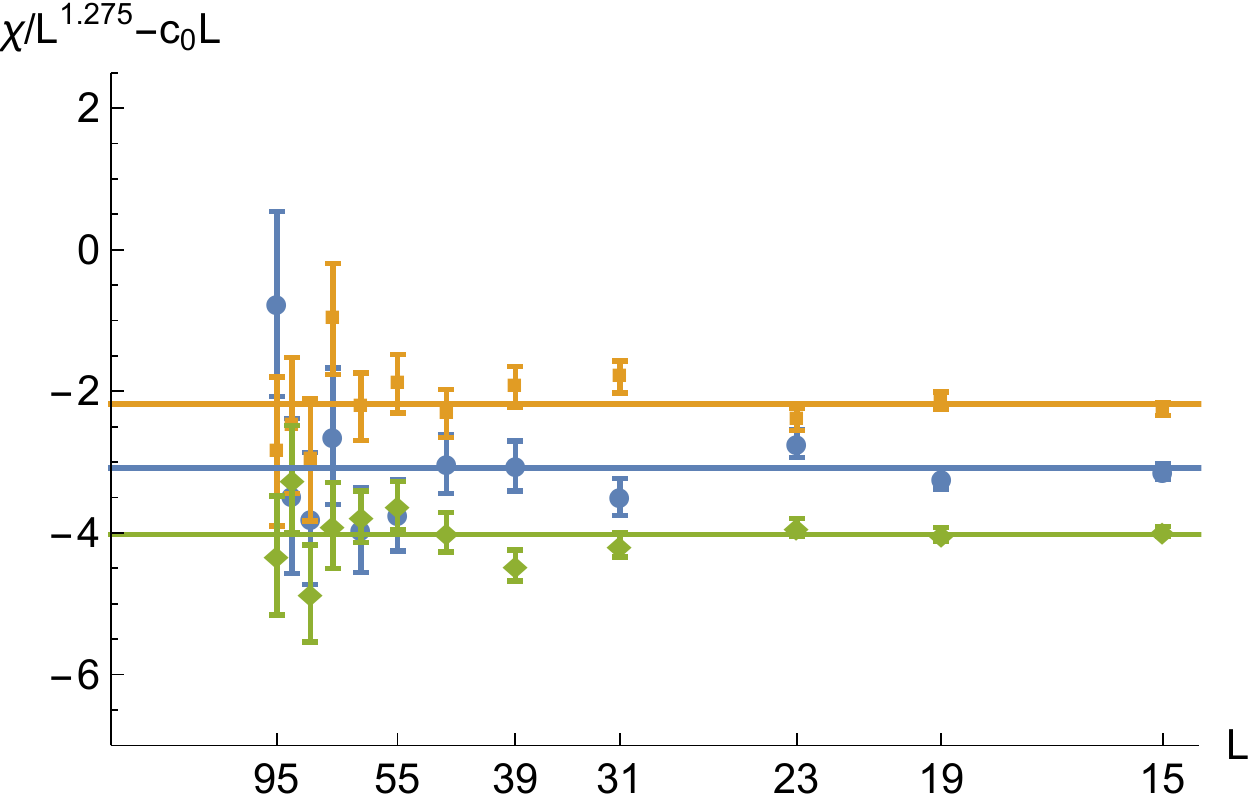}
  \caption{\label{fig:chi3a}(Colour on-line) $L(\chi/L^{2.275}-c_0)$
    versus $1/L$ for $L=15$, $19$, $23$, $31$, $39$, $47$, $55$, $63$,
    $71$, $79$, $87$ and $95$. Cases are $B_1$ (blue), $B_2$ (orange)
    and $B_4$ (green).  Values of $c_0$ are, respectively, $2.45$,
    $1.90$, $1.50$.  Constant lines are, respectively, $y=-3.1$,
    $y=-2.2$, $y=-4.0$. See Fig.~\ref{fig:chi3}}
\end{figure}

In Figs.~\ref{fig:chi2}, \ref{fig:chi3} and \ref{fig:chi4} we show the
normalised susceptibility $\chi/L^a$ versus $1/L$ together with the
fitted lines $y=c_0+c_1x$ from which we read the asymptotic value
$c_0$. This very simple rule appears quite sufficient and we see no
prescense of any higher-order correction terms. In fact, plotting
$c_1= L(\chi/L^a-c_0)$ versus $1/L$ is effectively constant (modulo
noise, increasing with $L$). We show here only the case of $B_r$ in
Fig.~\ref{fig:chi3a} but the other cases results in quite similar
plots. 

\section{Magnetisation distribution}
We will here make an attempt to describe the distribution of the
magnetisation.  Beginning with the kurtosis $\mkurt$ for PBC we expect
it to take the asymptotic value $\Gamma(1/4)^4/(8\pi^2)\approx
2.1884$~\cite{brezin:85} and for FBC we expect it to be $3$, as is
characteristic for a normal distribution.

In general the mean-field density function~\cite{binder:85,brezin:85}
\begin{equation}\label{fdef}
  f(x)=c_0\exp(-c_2x^2-c_4x^4)
\end{equation}
fits these distributions very well, except for very small systems
where a correction factor is needed. Note that the case $c_2=0$ gives
the kurtosis $2.1884$ mentioned above for all $c_4>0$. The
distribution then is unimodal when $c_2>0$ giving $\mkurt>2.1884$ and
a bimodal distribution when $c_2<0$ corresponding to
$\mkurt<2.1884$. The function $f(x)$ can now be determined from the
variance and the kurtosis. We will here only show the standardised
form (variance $1$).

\begin{figure}
  \includegraphics[width=0.483\textwidth]{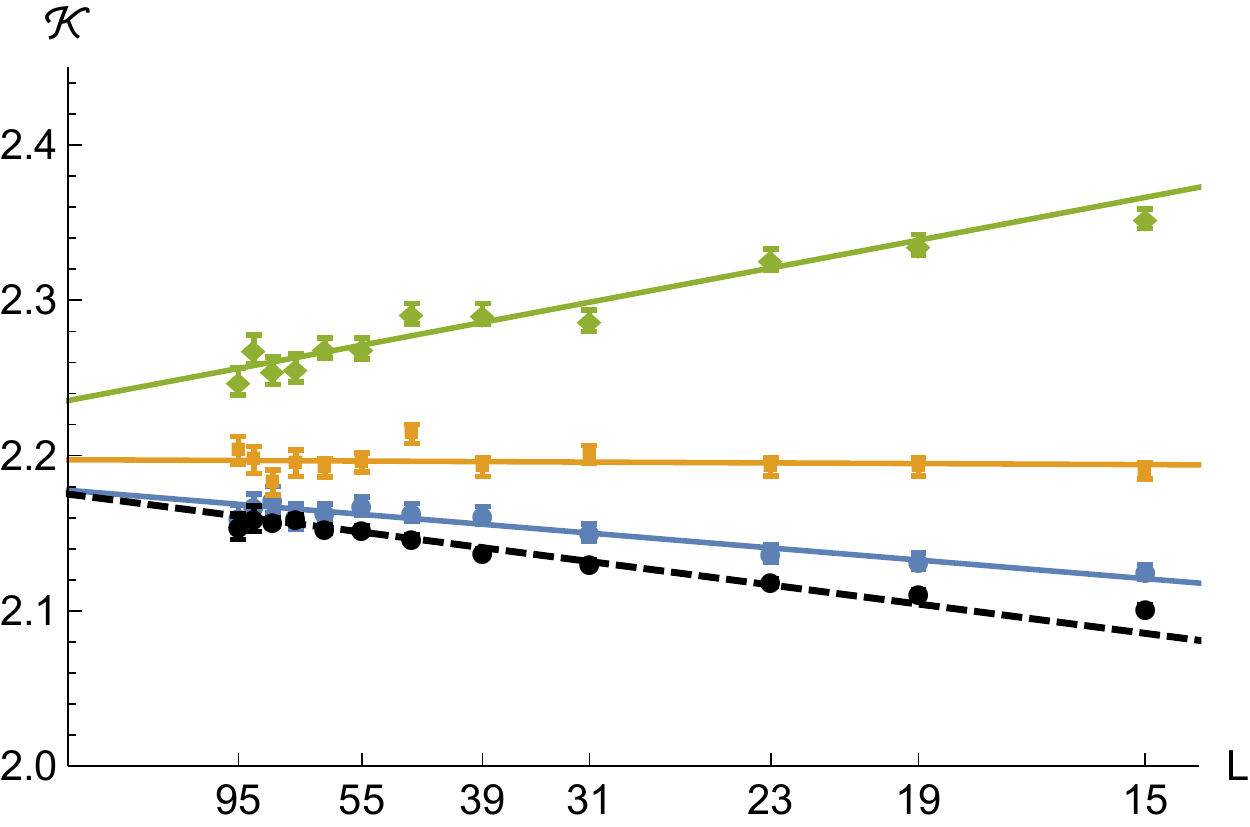}
  \caption{\label{fig:mk02}(Colour on-line) Magnetisation kurtosis
    $\mkurt(\beta_c)$ versus $L$ for $L=15$, $19$, $23$, $31$, $39$,
    $47$, $55$, $63$, $71$, $79$, $87$ and $95$. Cases are, PBC
    (black, dashed curve), $A_1$ (blue), $A_2$ (orange) and $A_4$
    (green). Lines fitted to $L\ge 19$ give asymptotic values,
    respectively, $2.18(1)$, $2.18(1)$, $2.20(1)$, $2.24(1)$. }
\end{figure}

The kurtosis plotted in Fig.~\ref{fig:mk02} demonstrate an interesting
feature with regard to the distribution shape. Since the cases PBC and
$A_1$ all have $\mkurt < 2.1884$ they are thus bimodal but converge to
a ``flat'' distribution ($c_2=0$). However, in the case of $A_2$ the
kurtosis is almost constant $\approx 2.2$ for all $L$, just barely
unimodal. For $A_4$ we are safely into unimodal territory though.

\begin{figure}
  \includegraphics[width=0.483\textwidth]{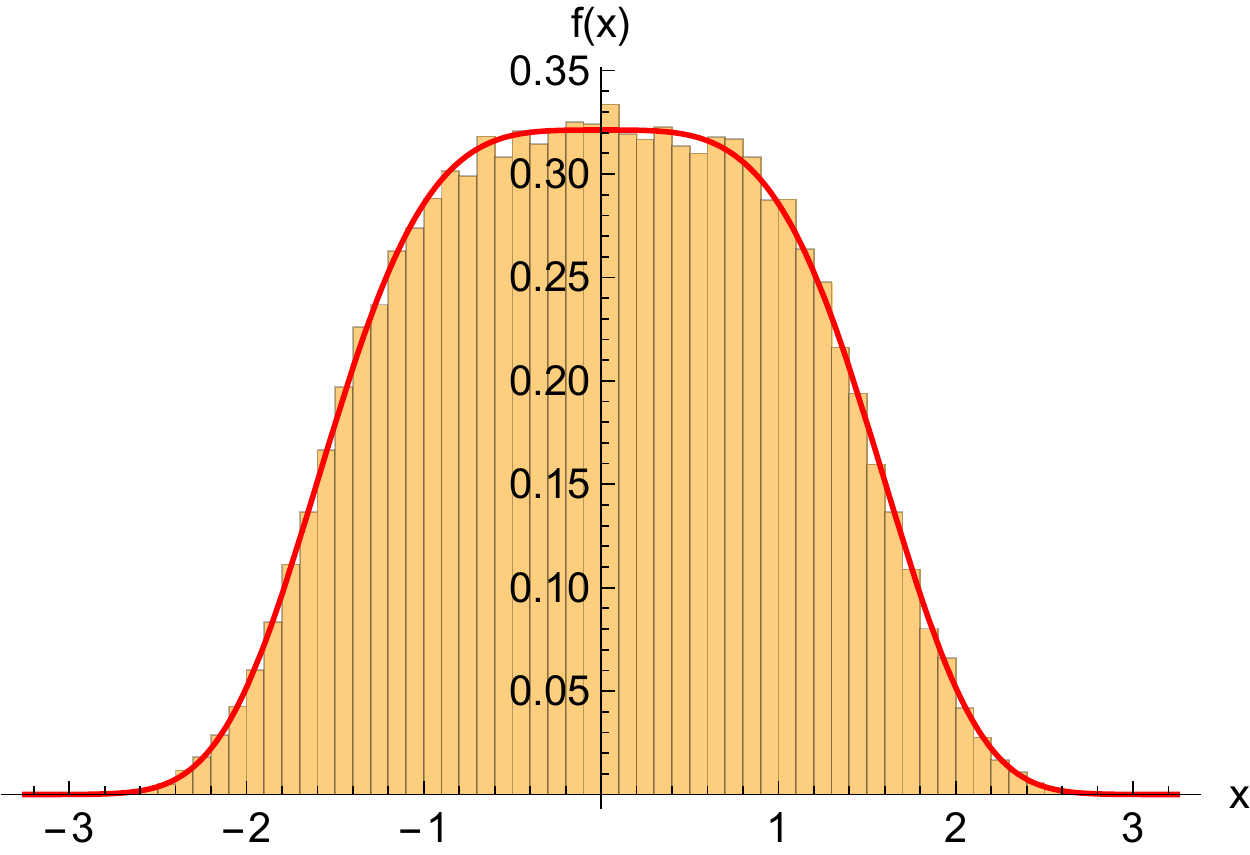}
  \caption{\label{fig:d2}(Colour on-line) Standardised magnetisation
    distribution density for $A_2$-case, $L=63$. Kurtosis $\mkurt =
    2.192(6)$ gives $f(x)=0.3213 \exp(-0.00445x^2-0.113x^4)$ (red
    curve) where $x=M/\sigma$.  A Pearson goodness-of-fit test is
    passed with $p$-value $0.61$.}
\end{figure}

In Fig.~\ref{fig:d2} we show an example of a standardised distribution
in the $A_2$-case for $L=63$.  Based on the kurtosis we find the
density function $f(x)$ numerically.  Here $\mkurt=2.1922$ which gives
$f(x)=0.3213\exp(-0.004541x^2-0.1130x^4)$, where $x=M/\sigma$. A
Pearson goodness-of-fit test is now used to see if we should reject
the hypothesis that $f(x)$ fits the magnetisation distribution. In
fact we find $\chi^2/\mathrm{dof}\approx 0.96$ for $\mathrm{dof}=200$
and a $p$-value of $0.61$ so we choose not to reject this.

This was repeated for the other cases (three $A_r$) and sizes (twelve
$L$) as well, using $\lceil 2n_s^{2/5}\rceil$ equiprobable bins
(Mathematica's default) for $n_s$ samples. Median $p$-value over these
$36$ instances is $0.50$ with interquartile range $0.47$ and all were
larger than $0.05$.  The hypothesis that the distribution is described
by Eq.~\eqref{fdef} is thus not rejected for $A_r$ and $L\ge 15$.

\begin{figure}
  \includegraphics[width=0.483\textwidth]{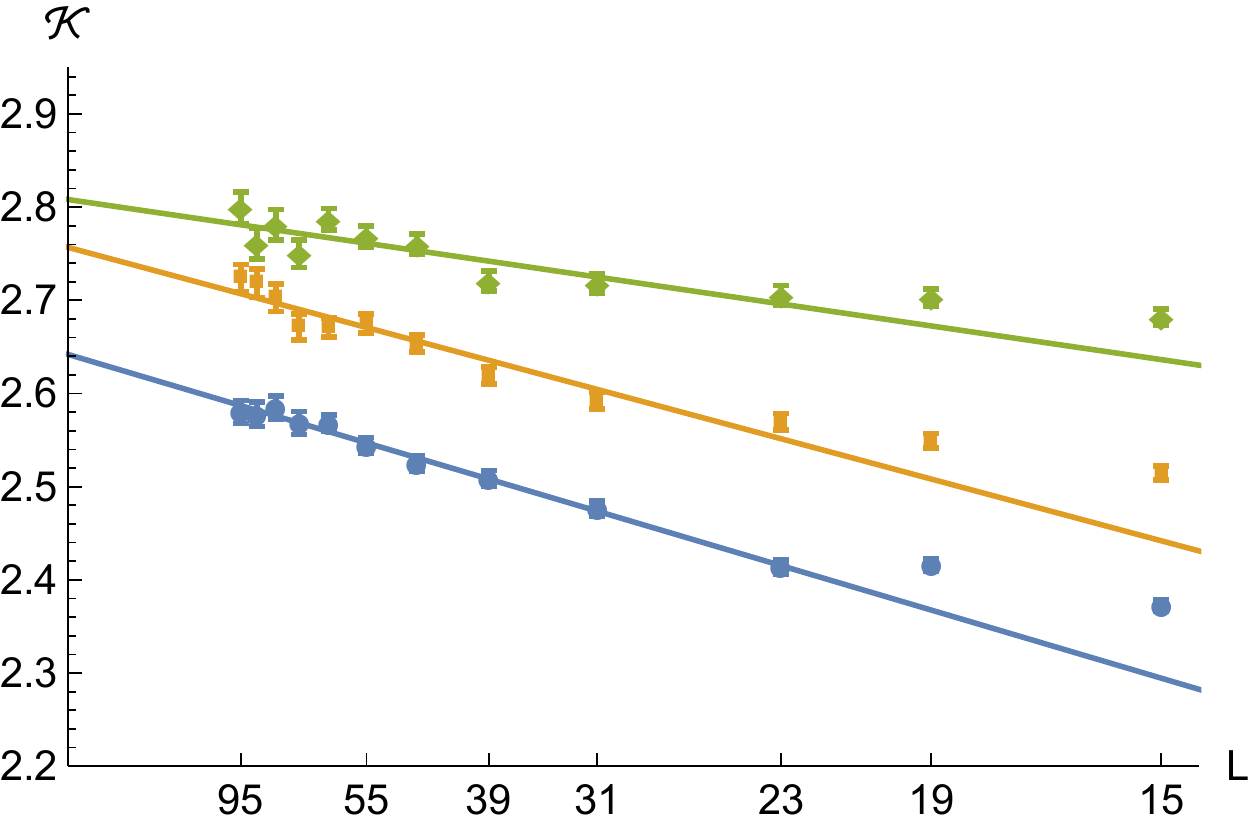}
  \caption{\label{fig:mk3}(Colour on-line) Magnetisation kurtosis
    $\mkurt(\beta_c)$ versus $L$ for $L=15$, $19$, $23$, $31$, $39$,
    $47$, $55$, $63$, $71$, $79$, $87$ and $95$. Cases are $
    B_1$ (blue), $B_2$ (orange) and $B_4$ (green). Lines fitted to
    $L\ge 23$ give asymptotic values, respectively, $2.64(1)$,
    $2.76(1)$, $2.81(1)$. }
\end{figure}

\begin{figure}
  \includegraphics[width=0.483\textwidth]{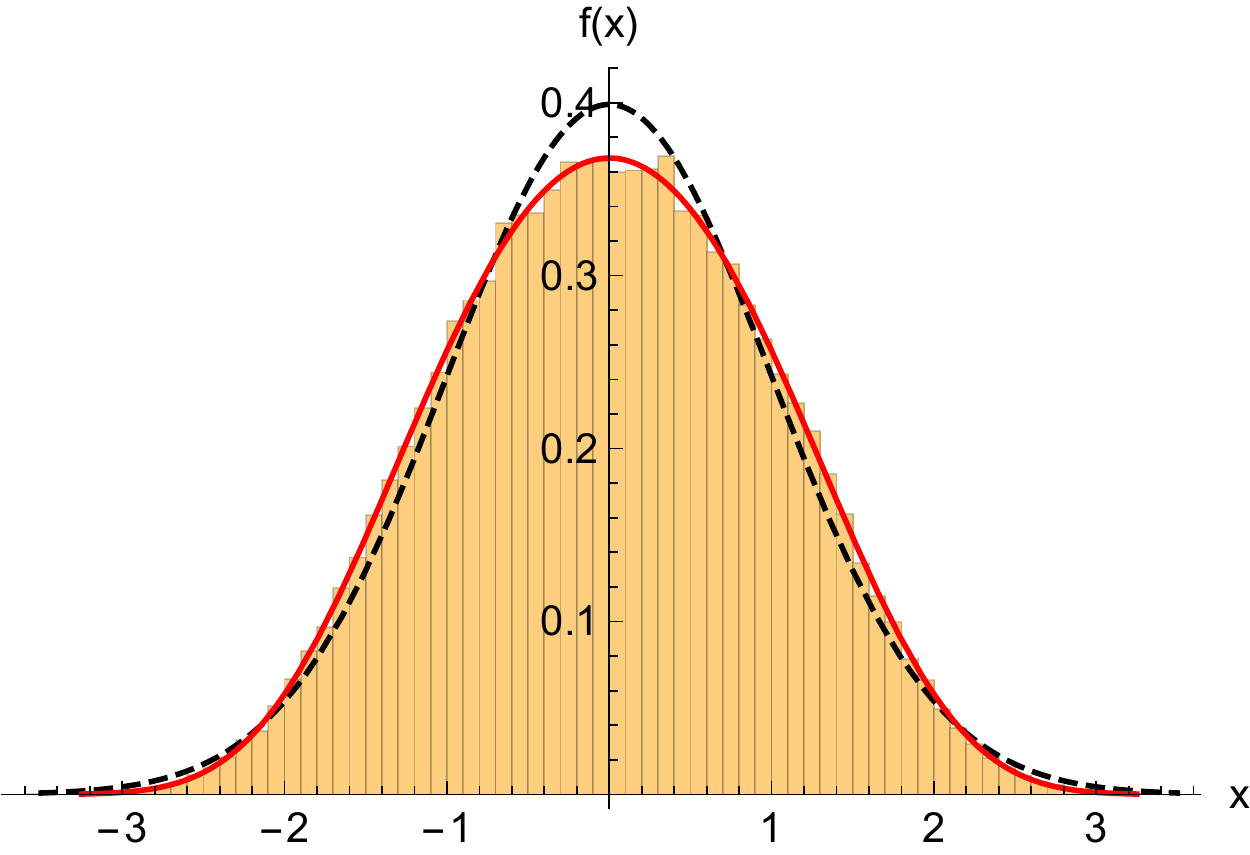}
  \caption{\label{fig:d3}(Colour on-line) Standardised magnetisation
    distribution density for $L=63$ with edges $
    B_1$. Kurtosis $\mkurt = 2.568(9)$ gives $f(x)=0.3680
    \exp(-0.327x^2-0.0338x^4)$ (red curve) where $x=M/\sigma$. Dashed
    black curve is density of a normal distribution. A Pearson
    goodness-of-fit test is passed with $p$-value $0.80$.}
\end{figure}

Moving on to the case $B_r$ we show the magnetisation kurtosis in
Fig.~\ref{fig:mk3} and it does not appear to converge to $3$ for any
of these cases. The distributions still fit Eq.~\eqref{fdef}
though. The Pearson test gave the median $p$-value $0.71$ and an
interquartile range of $0.48$ and only one of the 36 instances ($r=1$,
$L=55$) gave $p<0.05$. Since $5$\% of the instances will fail even if
the hypothesis is true we find this quite normal. We thus do not
reject that Eq.~\eqref{fdef} fits these distributions for $
B_r$ and $L\ge 15$. In Fig.~\ref{fig:d3} we show an example of this
distribution and $f(x)$ for $B_1$ and $L=63$.

For the $C_r$-case the kurtosis is close to $3$ (modulo noise) in
almost all instances. Only for $r=1$ do we see a weak trend with
values clearly distinct from $3$ for the smallest $L$ (no figure).  We
tested the hypothesis that the distribution of the magnetisation
samples are Gaussian for all $r$ and $L\ge 23$. Indeed, of these $50$
instances only $4$ ($8$\%) fail ($p$-value less than $0.05$) which is
to be expected. The median $p$-value over the instances was $0.46$
with interquartile range $0.45$. Thus we do not reject the hypothesis.

\section{Scaling of energy quantities}
Let us first discuss the finite-size scaling of the internal energy
$\erg$.  Across the different cases the data seem to agree on the
common limit energy $\erg(\beta_c)=0.6756(1)$. For PBC a very simple
scaling rule $\erg_L=c_0+c_1/L^{5/2}$ is sufficient. For $A_r$ we
suggest $\erg_L=c_0+c_1/L^{5/2}+c_2/L^5$, though the choice of the
second exponent is uncertain. 

\begin{figure}
  \includegraphics[width=0.483\textwidth]{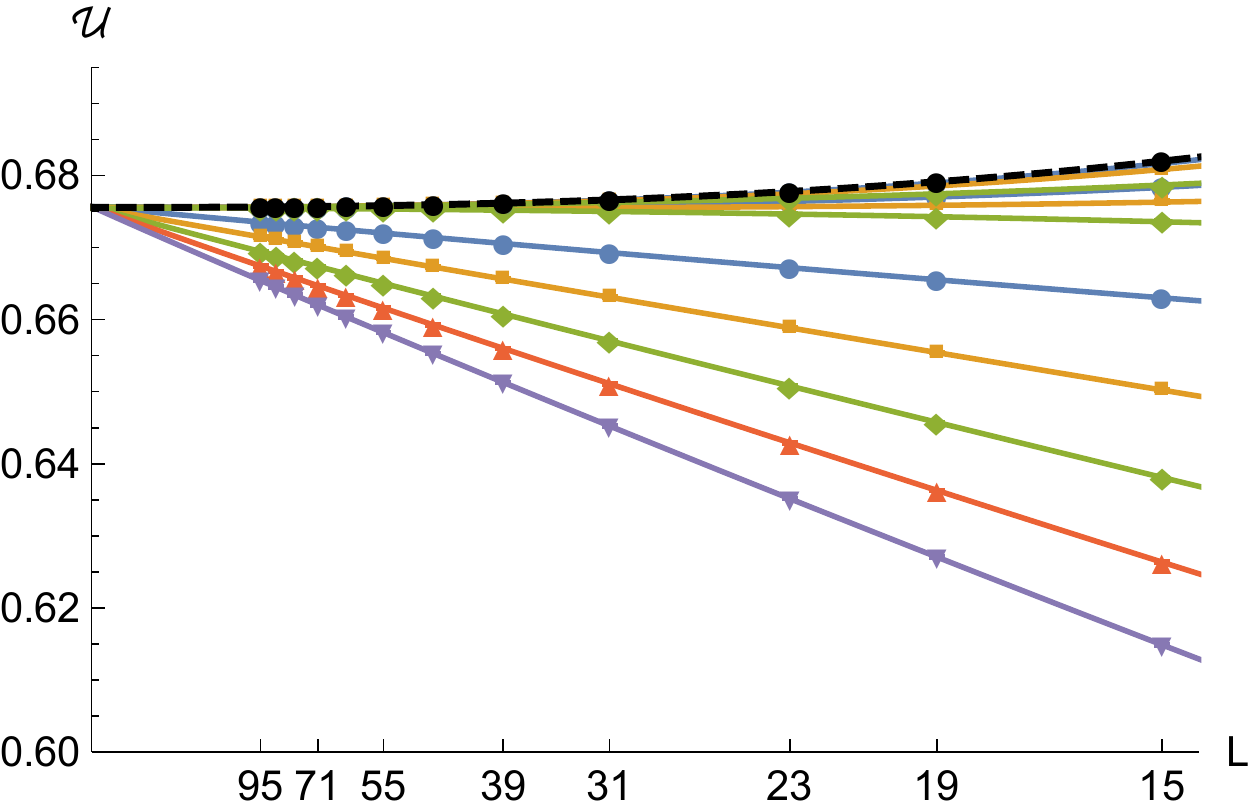}
  \caption{\label{fig:u}(Colour on-line) Energy $\erg$ versus $L$ for
    $L=15$, $19$, $23$, $31$, $39$, $47$, $55$, $63$, $71$, $79$, $87$
    and $95$. Boundary conditions are (downwards) PBC (black, dashed
    curve), $A_r$ ($r=1,2,4$), $B_r$ ($r=1,2,4$) and $C_r$
    ($r=1,\ldots,5$) The fitted curves (see text) have the common
    limit $0.6756(1)$, Error bars are shown but smaller than the
    points.}
\end{figure}

At the other end of the spectrum, for FBC and $C_r$, the rule
$\erg_L=c_0+c_1/L+c_2/L^{3/2}$ gives stable scaling, also used in
Ref.~\cite{lundow:14}. For $B_r$ the rule
$\erg_L=c_0+c_1/L+c_2/L^{5/2}$ gives stable behaviour when deleting
points. These scaling rules gives the limit value $c_0=0.6756(1)$
above but we seem unable to provide any more digits. Needless to say,
we have no theory-based support for these scaling rules.  In
Fig.~\ref{fig:u} we plot the energy versus $1/L$ for all cases and
sizes together with fitted curves as just described.


\begin{figure}
  \includegraphics[width=0.483\textwidth]{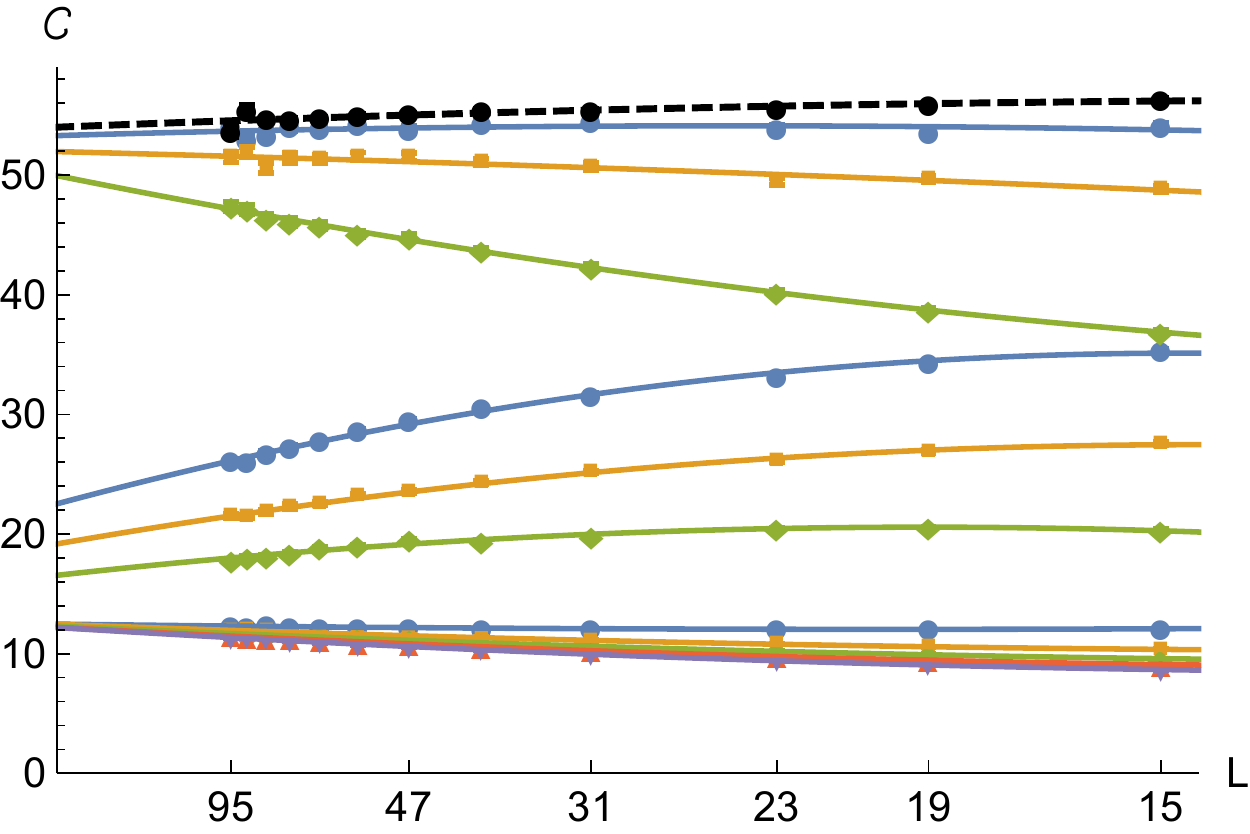}
  \caption{\label{fig:c}(Colour on-line) Specific heat $\heat$ versus
    $L$ for $L=15$, $19$, $23$, $31$, $39$, $47$, $55$, $63$, $71$,
    $79$, $87$ and $95$. Boundary conditions are (downwards at
    $y$-axis) PBC (black, dashed curve), $A_r$ ($r=1,2,4$), $B_r$
    ($r=1,2,4$) and $C_r$ ($r=1,\ldots,5$). Limit values range from
    $54$ for PBC to $12.5$ for FBC (see text).  Error bars are shown
    but smaller than the points.}
\end{figure}

Since the specific heat is bounded for 5D systems~\cite{sokal:79} one
might expect its finite-size scaling to be similar to that of the
energy.  Unfortunately the scaling rules above do not seem to apply to
the specific heat. In Fig.~\ref{fig:c} we therefore plot the specific
heat together with fitted 2nd degree polynomials which at least provides
rough estimates of the limit values.

For PBC we estimate the limit $\heat(\beta_c)\approx 54.0(5)$
(marginally less than in Ref.~\cite{lundow:15}). In the $A_r$-case we
obtain the limits $53.0(5)$, $52.0(5)$, $50.0(5)$ for $r=1,2,4$,
respectively. The $B_r$-case gives $22.5(5)$, $19.0(5)$, $16.5(5)$ for
$r=1,2,4$, respectively.  Finally, the $C_r$-case resulted in the
(probably) common estimate $12.5(5)$, though the noise is larger than
the differences between the estimated limits.

In Ref.~\cite{lundow:14} only one correction term was used with
exponent $1/3$ for FBC, so that $\heat_L=c_0+c_1/L^{1/3}$. All
$C_r$-instances are indeed well-fitted by this elegant rule, but this
would lead to a limit of $14.7$ for $C_5$ (FBC), larger than the
resulting limit $12.6$ for $C_1$. Hence, we hesitate to use this
simple rule since all data suggest that the specific heat should
decrease when deleting boundary edges, just like the energy does in
Fig.~\ref{fig:u}.

\begin{figure}
  \includegraphics[width=0.483\textwidth]{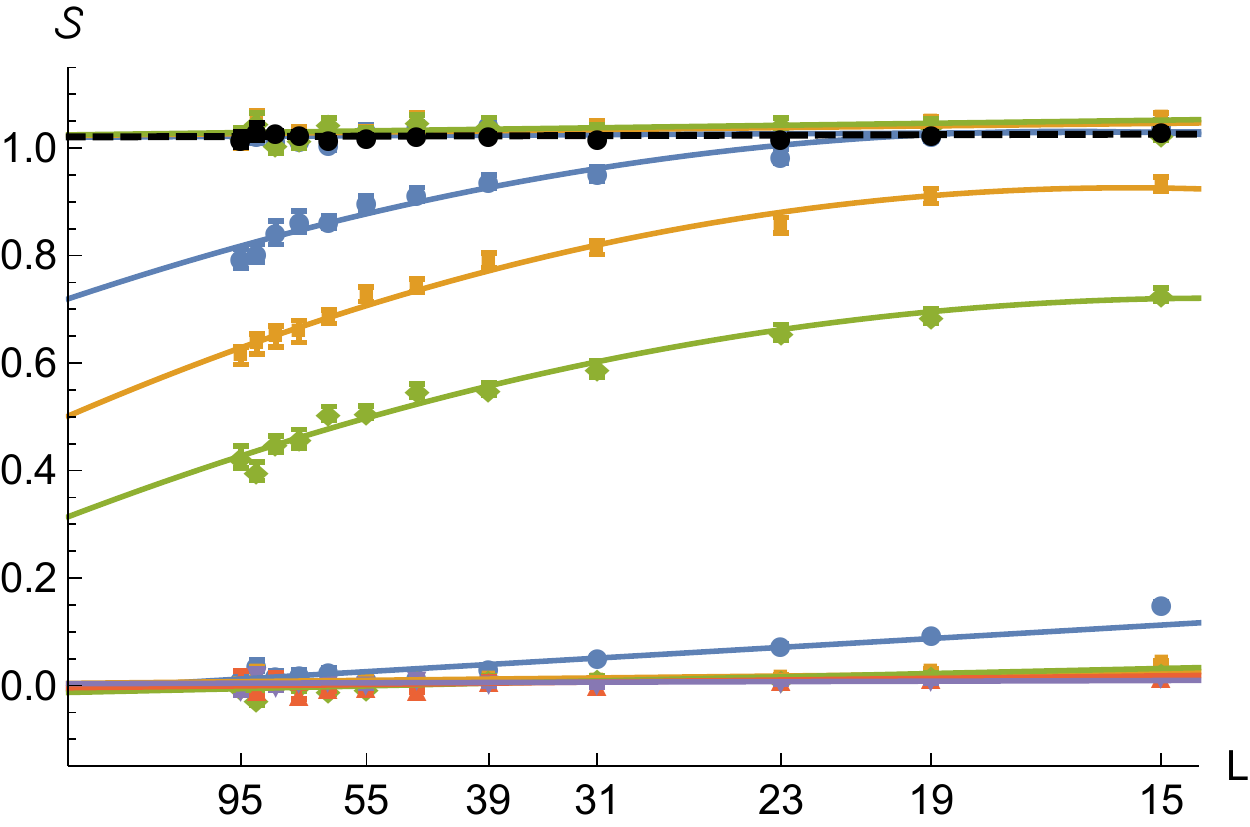}
  \caption{\label{fig:es}(Colour on-line) Energy skewness $\eskew$
    versus $L$ for $L=15$, $19$, $23$, $31$, $39$, $47$, $55$, $63$,
    $71$, $79$, $87$ and $95$. Boundary conditions are: at the top,
    PBC (black, dashed curve) and $A_r$ ($r=1,2,4$,
    indistinguishable); middle three, $B_r$ ($r=1,2,4$ downwards); at
    the bottom, $C_r$ ($r=1,\ldots,5$, with $r=1$ deviating). Limit
    values range from $1.02$ for PBC and $A_r$ to $0$ for FBC (see
    text for details).  Error bars are shown but often smaller than
    the points.}
\end{figure}

In Fig.~\ref{fig:es} we see the effect that the boundary has on the
skewness of the energy distribution. At the top of the figure we see
that PBC and $A_r$ are almost indistinguishable and they all agree on
a common limit value of $1.023(5)$ based on fitted lines. For PBC and
$A_1$ the skewness is effectively constant (modulo noise) over
$L$. There is only a very small size-dependence for $A_2$ and $A_4$.

For $B_r$ the skewness has clearly separated itself from PBC. Also,
the size-dependence becomes clearly nonlinear. Fitting 2nd degree
polynomials to the points we estimate the limits $0.72(1)$, $0.50(1)$
and $0.31(1)$ for $r=1,2,4$ respectively (error bars from deleting one
point in the fit).  In the $C_r$-case the skewness is practically zero
for all $L$ for $r=2,3,4,5$ but there is a distinct (almost) linear
size-dependence for $r=1$, becoming effectively zero for $L\ge 47$.

\begin{figure}
  \includegraphics[width=0.483\textwidth]{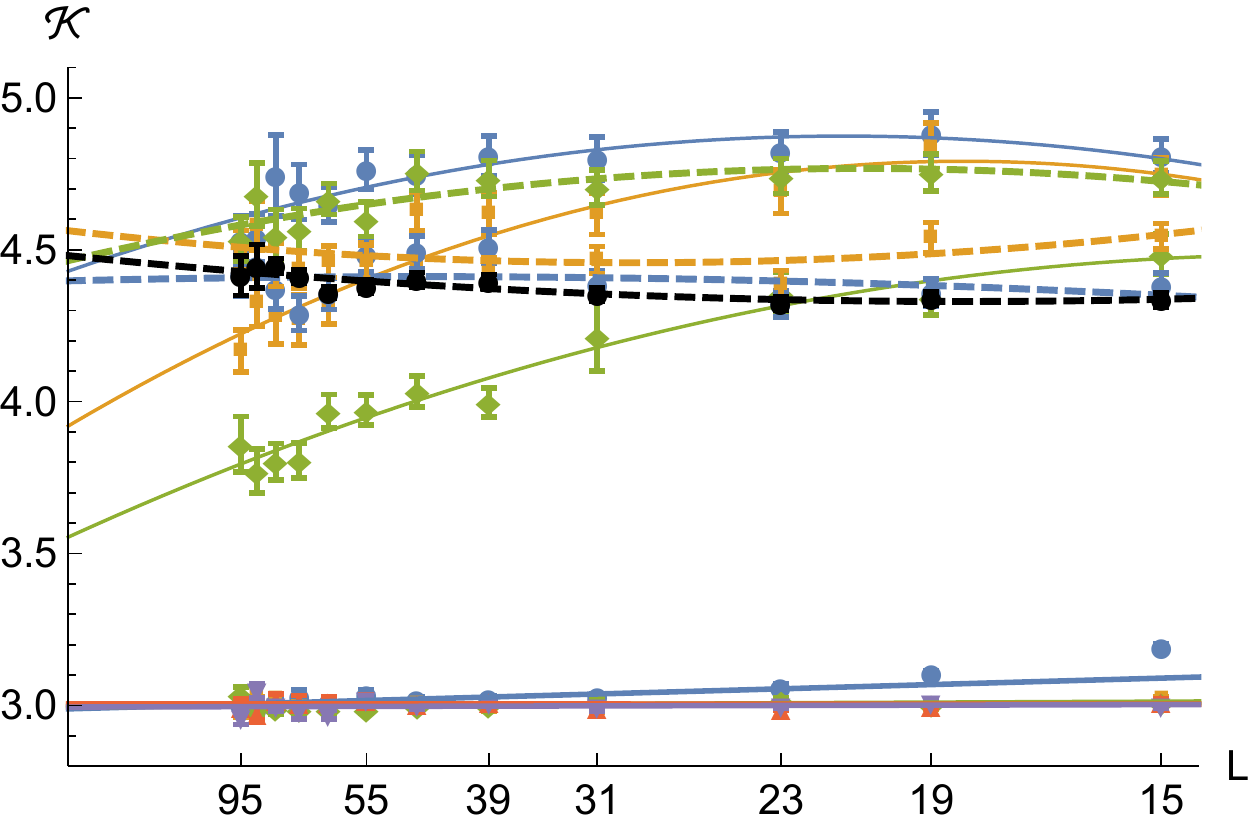}
  \caption{\label{fig:ek}(Colour on-line) Energy kurtosis $\ekurt$
    versus $L$ for $L=15$, $19$, $23$, $31$, $39$, $47$, $55$, $63$,
    $71$, $79$, $87$ and $95$. Boundary conditions are: at the top,
    PBC (black, dashed curve) and $A_r$ ($r=1,2,4$, (blue, orange,
    green; dashed curves); $B_r$ ($r=1,2,4$ blue, orange, green; solid
    curves); at the bottom, $C_r$ ($r=1,\ldots,5$, with $r=1$
    deviating). See text for limit values.  Error bars are shown but
    sometimes smaller than the points.}
\end{figure}

\begin{figure}
  \includegraphics[width=0.483\textwidth]{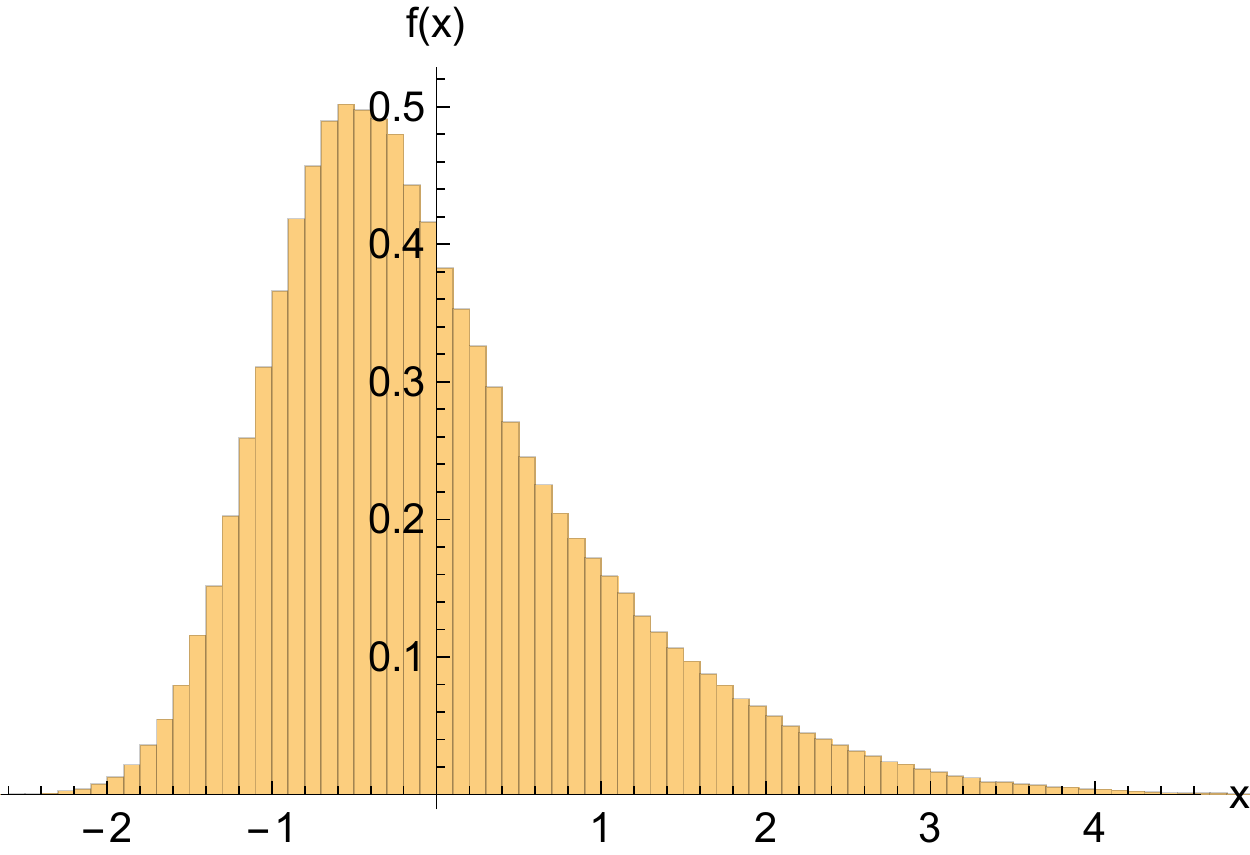}
  \caption{\label{fig:edist}(Colour on-line) Standardised energy
    distribution density for PBC, $L=79$, $n_s=500000$. This instance
    has energy $\erg=0.6756$, specific heat $\heat=54.7$, skewness
    $\eskew=1.03$ and kurtosis $\ekurt=4.44$. Note that
    $x=(E-\mean{E})/\sqrt{\var{E}}$.}
\end{figure}

The kurtosis $\ekurt$ of the energy distribution is shown in
Fig.~\ref{fig:ek}. At this point the pattern is clear.  PBC and $A_r$
behave similarly (dashed curves in the figure), though $A_4$ stands
out, with limit values $4.48(1)$ (PBC), $4.39(4)$ ($r=1$), $4.56(3)$
($r=2$) and $4.46(3)$ ($r=4$), with error bars estimated by removing
one point at a time from a fitted 2nd degree polynomial. Since the
error bars are larger for $A_r$ they could in fact have a common limit
value $4.48$.  In Fig.~\ref{fig:edist} we show the energy distribution
for PBC ($L=79$), which is quite similar to that of $A_r$.

For $B_r$ (solid curves in figure) we estimate the limits $4.43(3)$
($r=1$), $3.93(3)$ ($r=2$) and $3.55(3)$ ($r=4$) from fitted 2nd
degree polynomials. Possibly the $r=1$-case has the same limit value
as PBC. The $C_r$-case all favour the estimate $3.00(1)$ by way of
fitted lines, but there is a small size dependence for $r=1$.

We performed a Pearson goodness-of-fit test to compare the energy
distributions of $C_r$ to a Gaussian distribution. For $r=2,\ldots,5$,
this test is passed for all $L\ge 31$, giving a median $p$-value of
$0.47$ and interquartile range $0.34$ over the 36 instances. However,
for $r=1$ we need $L\ge 63$ to pass the test on the 5\%-level..

\section{Conclusion}
We have investigated several scenarios of deleting boundary edges from
a PBC-system, where an FBC-system corresponds to deleting $5L^4$
edges.  For example, what happens to the finite-size scaling of the
susceptibility when we delete all boundary edges in only one
direction, i.e., $L^4$ edges?  Remarkably, we find that this is enough
to switch the scaling behaviour to that associated with FBC. The
distribution of magnetisations, essentially distributed as
$c_0\exp(-c_4x^4)$ for PBC, has also switched to Gaussian for $L\ge
23$, to the extent that it passes a Pearson goodness-of-fit test.

Deleting just $L^2$ boundary edges does not change the scaling
behaviour significantly from that of PBC though it does increase the
kurtosis slightly, making the distribution of magnetisations unimodal.
However, deleting $L^3$ boundary edges changes the susceptibility
scaling to something strictly inbetween PBC and FBC, we estimate that
$\chi\propto L^{2.275}$. 

We also note that all magnetisation distributions are well-fitted by
the simple formula in Eq.~\eqref{fdef}, passing a Pearson test to the
expected degree.

The energy-related quantities also change when deleting boundary
edges, though correction-to-scaling terms give less than clear scaling
rules. We suggest that deleting $rL^4$ edges ($C_r$-case) may all give
the same specific heat limit value of $12.5(5)$, for
$r=1,\ldots,5$. Deleting just $rL^2$ edges, and for that matter,
$rL^3$ edges, gives us specific heat values inbetween PBC and FBC.

The energy skewness is essentially the same for PBC and when deleting
$rL^2$ edges, $\eskew\to 1.02$. Deleting $rL^4$ edges gives
$\eskew\to 0$. Deleting $rL^3$ edges gives limit values inbetween
these two.

The energy kurtosis for PBC and when deleting $rL^2$ edges may have
the same limit value, say $4.5$, but there is too much noise to say
with any certainty. Deleting $rL^4$ puts the kurtosis close to 3 and,
as we may expect, a Pearson test suggests the energy distribution is
essentially Gaussian in this case, if $L$ is large enough.  However,
deleting $rL^3$ edges puts the kurtosis somewhere inbetween, possibly
$r=1$ may give the same limit as PBC.

One may well wonder how this generalises to higher dimensions.  For
example, starting with a 6D system with periodic boundary conditions
we expect the finite-size scaling $\chi\propto L^3$. If we delete the
$L^5$ boundary edges along one direction, is this enough to change the
scaling to $\chi\propto L^2$?  What happens when we delete $L^4$,
$L^3$, $L^2$ edges?

\begin{acknowledgments}
  The computations were performed on resources provided by the Swedish
  National Infrastructure for Computing (SNIC) at Chalmers Centre for
  Computational Science and Engineering (C3SE).
\end{acknowledgments}


\begin{thebibliography}{15}
\expandafter\ifx\csname natexlab\endcsname\relax\def\natexlab#1{#1}\fi
\expandafter\ifx\csname bibnamefont\endcsname\relax
  \def\bibnamefont#1{#1}\fi
\expandafter\ifx\csname bibfnamefont\endcsname\relax
  \def\bibfnamefont#1{#1}\fi
\expandafter\ifx\csname citenamefont\endcsname\relax
  \def\citenamefont#1{#1}\fi
\expandafter\ifx\csname url\endcsname\relax
  \def\url#1{\texttt{#1}}\fi
\expandafter\ifx\csname urlprefix\endcsname\relax\def\urlprefix{URL }\fi
\providecommand{\bibinfo}[2]{#2}
\providecommand{\eprint}[2][]{\url{#2}}

\bibitem[{\citenamefont{Brezin and Zinn-Justin}(1985)}]{brezin:85}
\bibinfo{author}{\bibfnamefont{E.}~\bibnamefont{Brezin}} \bibnamefont{and}
  \bibinfo{author}{\bibfnamefont{J.}~\bibnamefont{Zinn-Justin}},
  \bibinfo{journal}{Nucl. Phys. B} \textbf{\bibinfo{volume}{257}},
  \bibinfo{pages}{867} (\bibinfo{year}{1985}).

\bibitem[{\citenamefont{Binder et~al.}(1985)\citenamefont{Binder, Nauenberg,
  Privman, and Young}}]{binder:85}
\bibinfo{author}{\bibfnamefont{K.}~\bibnamefont{Binder}},
  \bibinfo{author}{\bibfnamefont{M.}~\bibnamefont{Nauenberg}},
  \bibinfo{author}{\bibfnamefont{V.}~\bibnamefont{Privman}}, \bibnamefont{and}
  \bibinfo{author}{\bibfnamefont{A.~P.} \bibnamefont{Young}},
  \bibinfo{journal}{Phys. Rev. B} \textbf{\bibinfo{volume}{31}},
  \bibinfo{pages}{1498} (\bibinfo{year}{1985}).

\bibitem[{\citenamefont{Bl\"ote and Luijten}(1997)}]{blote:97}
\bibinfo{author}{\bibfnamefont{H.~W.~J.} \bibnamefont{Bl\"ote}}
  \bibnamefont{and} \bibinfo{author}{\bibfnamefont{E.}~\bibnamefont{Luijten}},
  \bibinfo{journal}{EPL (Europhysics Letters)} \textbf{\bibinfo{volume}{38}},
  \bibinfo{pages}{565} (\bibinfo{year}{1997}).

\bibitem[{\citenamefont{Luijten et~al.}(1999)\citenamefont{Luijten, Binder, and
  Bl\"ote}}]{luijten:99}
\bibinfo{author}{\bibfnamefont{E.}~\bibnamefont{Luijten}},
  \bibinfo{author}{\bibfnamefont{K.}~\bibnamefont{Binder}}, \bibnamefont{and}
  \bibinfo{author}{\bibfnamefont{H.~W.~J.} \bibnamefont{Bl\"ote}},
  \bibinfo{journal}{Eur. Phys. J. B} \textbf{\bibinfo{volume}{9}},
  \bibinfo{pages}{289} (\bibinfo{year}{1999}).

\bibitem[{\citenamefont{Rudnick et~al.}(1985)\citenamefont{Rudnick, Gaspari,
  and Privman}}]{rudnick:85}
\bibinfo{author}{\bibfnamefont{J.}~\bibnamefont{Rudnick}},
  \bibinfo{author}{\bibfnamefont{G.}~\bibnamefont{Gaspari}}, \bibnamefont{and}
  \bibinfo{author}{\bibfnamefont{V.}~\bibnamefont{Privman}},
  \bibinfo{journal}{Phys. Rev. B} \textbf{\bibinfo{volume}{32}},
  \bibinfo{pages}{7594} (\bibinfo{year}{1985}).

\bibitem[{\citenamefont{Camia et~al.}(2020)\citenamefont{Camia, Jiang, and
  Newman}}]{camia:20}
\bibinfo{author}{\bibfnamefont{F.}~\bibnamefont{Camia}},
  \bibinfo{author}{\bibfnamefont{J.}~\bibnamefont{Jiang}}, \bibnamefont{and}
  \bibinfo{author}{\bibfnamefont{C.~M.} \bibnamefont{Newman}},
  \bibinfo{journal}{ArXiv e-prints}  (\bibinfo{year}{2020}),
  \eprint{2011.02814}.

\bibitem[{\citenamefont{Lundow and Markstr\"om}(2011)}]{lundow:11}
\bibinfo{author}{\bibfnamefont{P.~H.} \bibnamefont{Lundow}} \bibnamefont{and}
  \bibinfo{author}{\bibfnamefont{K.}~\bibnamefont{Markstr\"om}},
  \bibinfo{journal}{Nucl. Phys. B} \textbf{\bibinfo{volume}{845}},
  \bibinfo{pages}{120 } (\bibinfo{year}{2011}).

\bibitem[{\citenamefont{Lundow and Markstr\"om}(2014)}]{lundow:14}
\bibinfo{author}{\bibfnamefont{P.~H.} \bibnamefont{Lundow}} \bibnamefont{and}
  \bibinfo{author}{\bibfnamefont{K.}~\bibnamefont{Markstr\"om}},
  \bibinfo{journal}{Nucl. Phys. B} \textbf{\bibinfo{volume}{889}},
  \bibinfo{pages}{249 } (\bibinfo{year}{2014}).

\bibitem[{\citenamefont{Berche et~al.}(2012)\citenamefont{Berche, Kenna, and
  Walter}}]{berche:12}
\bibinfo{author}{\bibfnamefont{B.}~\bibnamefont{Berche}},
  \bibinfo{author}{\bibfnamefont{R.}~\bibnamefont{Kenna}}, \bibnamefont{and}
  \bibinfo{author}{\bibfnamefont{J.-C.} \bibnamefont{Walter}},
  \bibinfo{journal}{Nucl. Phys. B} \textbf{\bibinfo{volume}{865}},
  \bibinfo{pages}{115 } (\bibinfo{year}{2012}).

\bibitem[{\citenamefont{Flores-Sola et~al.}(2016)\citenamefont{Flores-Sola,
  Berche, Kenna, and Weigel}}]{berche:16}
\bibinfo{author}{\bibfnamefont{E.}~\bibnamefont{Flores-Sola}},
  \bibinfo{author}{\bibfnamefont{B.}~\bibnamefont{Berche}},
  \bibinfo{author}{\bibfnamefont{R.}~\bibnamefont{Kenna}}, \bibnamefont{and}
  \bibinfo{author}{\bibfnamefont{M.}~\bibnamefont{Weigel}},
  \bibinfo{journal}{Phys. Rev. Lett.} \textbf{\bibinfo{volume}{116}},
  \bibinfo{pages}{115701} (\bibinfo{year}{2016}).

\bibitem[{\citenamefont{Wittmann and Young}(2014)}]{young:2014}
\bibinfo{author}{\bibfnamefont{M.}~\bibnamefont{Wittmann}} \bibnamefont{and}
  \bibinfo{author}{\bibfnamefont{A.~P.} \bibnamefont{Young}},
  \bibinfo{journal}{Phys. Rev. E} \textbf{\bibinfo{volume}{90}},
  \bibinfo{pages}{062137} (\bibinfo{year}{2014}).

\bibitem[{\citenamefont{Lundow and Markstr\"om}(2016)}]{lundow:16}
\bibinfo{author}{\bibfnamefont{P.~H.} \bibnamefont{Lundow}} \bibnamefont{and}
  \bibinfo{author}{\bibfnamefont{K.}~\bibnamefont{Markstr\"om}},
  \bibinfo{journal}{Nucl. Phys. B} \textbf{\bibinfo{volume}{911}},
  \bibinfo{pages}{163 } (\bibinfo{year}{2016}).

\bibitem[{\citenamefont{Wolff}(1989)}]{wolff:89}
\bibinfo{author}{\bibfnamefont{U.}~\bibnamefont{Wolff}},
  \bibinfo{journal}{Phys. Rev. Lett} \textbf{\bibinfo{volume}{62}},
  \bibinfo{pages}{361} (\bibinfo{year}{1989}).

\bibitem[{\citenamefont{Lundow and Markstr\"om}(2015)}]{lundow:15}
\bibinfo{author}{\bibfnamefont{P.~H.} \bibnamefont{Lundow}} \bibnamefont{and}
  \bibinfo{author}{\bibfnamefont{K.}~\bibnamefont{Markstr\"om}},
  \bibinfo{journal}{Nucl. Phys. B} \textbf{\bibinfo{volume}{895}},
  \bibinfo{pages}{305 } (\bibinfo{year}{2015}).

\bibitem[{\citenamefont{Sokal}(1979)}]{sokal:79}
\bibinfo{author}{\bibfnamefont{A.~D.} \bibnamefont{Sokal}},
  \bibinfo{journal}{Phys. Lett. A} \textbf{\bibinfo{volume}{71}},
  \bibinfo{pages}{451} (\bibinfo{year}{1979}).

\end{thebibliography}

\end{document}